\UseRawInputEncoding

\documentclass[conference]{IEEEtran}
\IEEEoverridecommandlockouts
\usepackage{cite}
\usepackage{amsmath,amssymb,amsfonts}
\usepackage{algorithmic}
\usepackage{graphicx}
\usepackage{textcomp}
\usepackage{xcolor}
\def\BibTeX{{\rm B\kern-.05em{\sc i\kern-.025em b}\kern-.08em
    T\kern-.1667em\lower.7ex\hbox{E}\kern-.125emX}}

\usepackage{tikz}

\usepackage{color,soul}
\usepackage[normalem]{ulem}
\usepackage{hyperref}
\usepackage{tabularx}
\usepackage{booktabs}
\usepackage{enumitem}
\usepackage{subcaption}
\usepackage{xcolor}
\usepackage{setspace}
\usepackage{graphicx}
\usepackage{amsmath}
\usepackage{amsfonts}
\usepackage{xspace}
\usepackage{multirow}
\usepackage{mathrsfs}
\usepackage{listings}
\usepackage[]{algorithm2e}

\newcommand*\circleb[1]{\tikz[baseline=(char.base)]{
            \node[shape=circle,draw,inner sep=0.5pt,fill=black,text=white, scale=0.85] (char) {#1};}}

\let\svthefootnote\thefootnote
\newcommand\freefootnote[1]{%
  \let\thefootnote\relax%
  \footnotetext{#1}%
  \let\thefootnote\svthefootnote%
}

\usepackage{etoolbox}

\newcommand{\siteName}{rMDC}
\newcommand{\name}{SkyBox}
 
\newcommand{\mdc}{MDC}

\newbool{IsPrintComment}
\booltrue{IsPrintComment}   

\newcommand{\shadi}[1]
{
	\ifbool{IsPrintComment}
	{
		{\color{magenta}(SN) #1}
	}
	\  
}
\newcommand{\todo}[1]
{
	\ifbool{IsPrintComment}
	{
		{\color{blue} #1}
	}
	\  
}

\newcommand{\maybe}[1]
{
	\ifbool{IsPrintComment}
	{
		{\color{orange}(Maybe Todo) #1}
	}
	\  
}

\newcommand{\shadiS}[1]
{
	\ifbool{IsPrintComment}
	{
		{\color{blue} #1}
	}
	{
		{\color{black} #1 }
	}
	\  
}

\newcommand{\anup}[1]
{
	\ifbool{IsPrintComment}
	{
		{\color{purple}(AA) #1}
	}
	\  
}

\newcommand{\anirudh}[1]
{
	\ifbool{IsPrintComment}
	{
		{\color{blue}(AB) #1}
	}
	\  
}

\newcommand{\TODO}[1]
{
	\ifbool{IsPrintComment}
	{
		{\color{red}todo: #1}
	}
	\  
}

\newcommand{\eg}{{e.g.}\xspace}

\definecolor{DrawBlue}{HTML}{DAE8FC}
\definecolor{DrawRed}{HTML}{F8CECC}
\definecolor{DrawPurple}{HTML}{E1D5E7}

\newcommand\vertarrowbox[3][6ex]{%
  \begin{array}[t]{@{}c@{}} #2 \\
  \left\uparrow\vcenter{\hrule height #1}\right.\kern-\nulldelimiterspace\\
  \makebox[0pt]{\scriptsize#3}
  \end{array}%
}

\begin{document}

\title{Exploring the Efficiency of Renewable Energy-based Modular Data Centers at Scale}
\date{}
\author{
    \IEEEauthorblockN{Jinghan Sun\IEEEauthorrefmark{2}*, Zibo Gong\IEEEauthorrefmark{2}*, Anup Agarwal\IEEEauthorrefmark{3}, Shadi Noghabi\IEEEauthorrefmark{4}, Ranveer Chandra\IEEEauthorrefmark{4}, Marc Snir\IEEEauthorrefmark{2}, Jian Huang\IEEEauthorrefmark{2}}
    \vspace{1ex}
    \IEEEauthorblockA{\IEEEauthorrefmark{2} University of Illinois Urbana-Champaign
    \\\{js39, zibog2, snir, jianh\}@illinois.edu}
    \vspace{0.5ex}
    \IEEEauthorblockA{\IEEEauthorrefmark{3}Carnegie Mellon University \hspace{20ex} \IEEEauthorrefmark{4}Microsoft Research  
    \\\hspace{8ex} anupa@andrew.cmu.edu \hspace{12ex} \{Shadi.Noghabi, ranveer\}@microsoft.com}
    \vspace{-3ex}
}

\maketitle
\begin{abstract}

Modular data centers (\mdc{s}) that can be placed right at the energy farms and powered mostly by renewable energy, are proven to be a flexible and effective approach to lowering the carbon footprint of data centers.
However, the main challenge of using renewable energy is the high variability of power produced, which implies 
large volatility in powering computing resources at \mdc{s}, and degraded application performance due to 
the task evictions and migrations. This causes challenges for platform operators to decide the \mdc{} deployment. 

To this end, we present~\name{}, a framework that employs a holistic and learning-based approach for platform operators to explore the efficient use of renewable energy with \mdc{} deployment across geographical regions. 
\name{} is driven by the insights based on our study of real-world power traces from a variety of renewable energy farms -- the predictable production of renewable energy and 
the complementary nature of energy production patterns across different renewable energy sources and locations. 
With these insights, \name{} first uses the coefficient of variation metric to select the qualified renewable farms, and 
proposes a subgraph identification algorithm to identify a set of farms with complementary energy production patterns. 
After that, \name{} enables smart workload placement and migrations to further tolerate the power variability. 
Our experiments with real power traces and datacenter workloads show that 
\name{} has the lowest carbon emissions in comparison with current \mdc{} deployment approaches. \name{} also minimizes the impact of the power variability on cloud virtual machines, enabling \siteName{s} a practical solution of efficiently using renewable 
energy.

\end{abstract}

\freefootnote{*Equal contribution.}
\pagestyle{plain}
\section{Introduction}
\label{sec:intro}

Data centers today consume more than 2\% of total U.S. power~\cite{datacentertotaconsumption}
and emit even more carbon than the aviation industry~\cite{lean-ict}. As a result, major datacenter vendors have to purchase carbon credits to temporarily offset the carbon impact of data centers~\cite{azurepledge,pledge,googlepledge}. 
However, this can only temporarily mitigate the carbon impact of data centers. 
To curb the carbon footprint of data centers at scale, a promising long-term solution is to rely more on renewable energy sources (\eg, solar 
and wind) as opposed to non-renewable sources (\eg, oil and gas), especially considering the recently reduced cost of renewable energy~\cite{renewablecost, carbonexplorer:asplos2023}. 
However, the key challenge of utilizing renewable energy is their variability across time and space. 
For instance, solar power production varies across time and the geographical locations. 

To manage the variability of renewable energy production, two common approaches have been explored.
The first approach is to transmit energy (a mix of renewable and non-renewable energy)
via transmission lines to data centers. This approach incurs significant monetary cost, 
additional carbon footprint, and increased complexity of power management due to the uncontrollable 
mix of renewable and non-renewable energy sources~\cite{carbonexplorer:asplos2023}. Specifically, 
about half of the cost is due to the power transmission and distribution~\cite{fares2017trends}.
As an alternative, batteries are used to store power and supply at a later time. However, the battery storage 
is expensive for large-scale datacenter deployment and they are minuscule in scale~\cite{battery_nrel}. For instance, the 
battery capacity in the US is only 0.4\% of the overall solar and wind capacity~\cite{storageus,renewableus}.

\begin{figure}[!t]
    \centering
    \includegraphics[width=0.8\columnwidth, trim={0 0 0cm 0cm}]{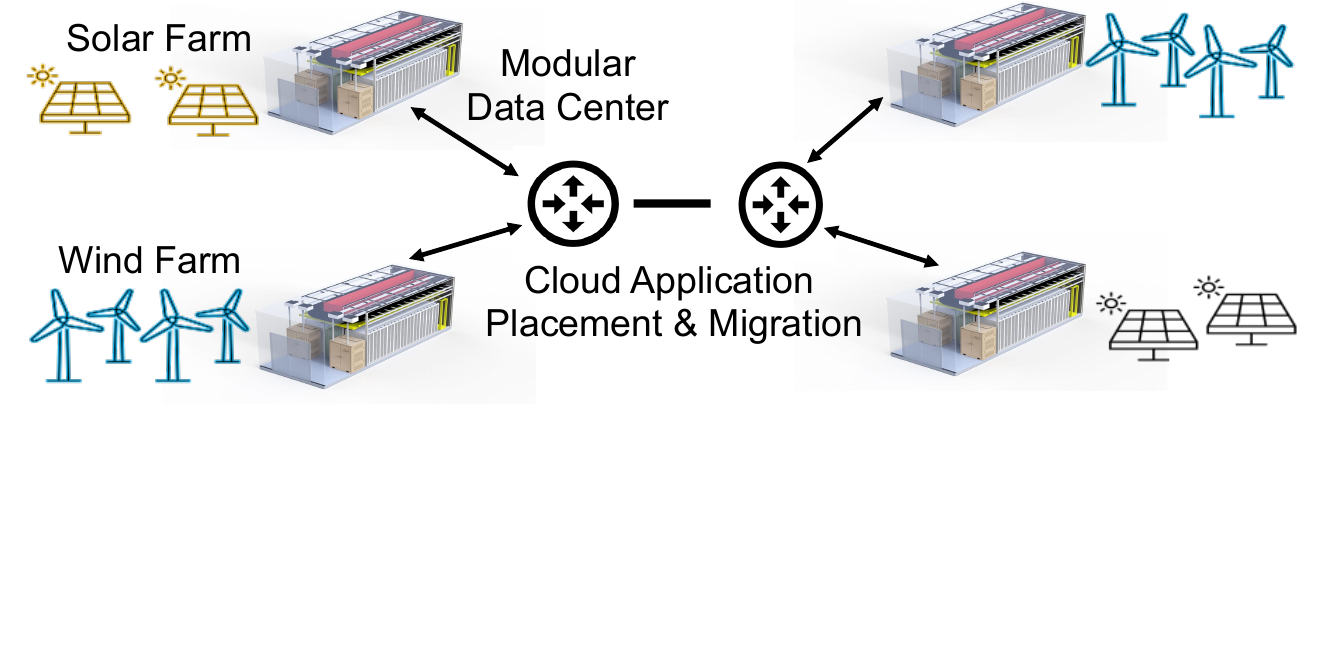}
    \vspace{-9.5ex}
	\caption{The target system architecture of \name{}. It facilitates the deployment of renewable energy-based modular data centers (\siteName{s})  
	across multiple geographical regions at scale. 
    }
    \label{fig:arch}
    \vspace{-3ex}
\end{figure}

To address these issues, both the academic and industry community~\cite{facebook:energy, ms:energy, zcc, parasol:asplos2013} proposed to co-locate data centers 
with renewable farms and powering data centers using renewable energy (Figure~\ref{fig:arch}). We define them as renewable energy-based modular data 
centers (\siteName{s}) in this paper. The \siteName{s} can alleviate the heavy use of batteries and power transmission lines, and have a great potential to achieve zero emissions. 
Moreover, modular data centers have low construction cost, and offer great mobility across different regions~\cite{parasol:asplos2013, inigo:micro2014, josep:icdcs2014}. We have also seen that 
many datacenter vendors have been planning to deploy \siteName{s} to meet the increasing 
demands on cloud computing from customers while achieving the carbon-free goal~\cite{ms:energy, facebook:energy, google:energy, aws:energy}. 

As we deploy modular data centers, we still suffer from the challenges caused by the variability of renewable energy produced 
in each renewable energy farm. To this end, we develop \name{}, a framework that uses a holistic and learning-based approach to enabling the 
efficient use of renewable energy with \siteName{} deployment at scale. 

With \name{}, we aim to answer three research questions: (1) where to deploy \siteName{s} across geographical regions? 
(2) how to maximize the efficiency of renewable energy sources across multiple \siteName{s}? and 
(3) how to enable smart application placement and migrations across \siteName{s} to minimize the performance impact of the 
power variability of renewables. By answering them, we wish \name{} will facilitate the deployment 
and operation of \siteName{s} at scale.

We drive the design of \name{} with three main observations based on our study of the distribution and production patterns of energy sources generated 
at more than 500 energy farms in total. 
We first observe that not all renewable farms are good candidates to 
place \siteName{s} as some farms consistently do not have enough stable power output. 
Second, the power production can be complementary across multiple geographically distributed energy farms, and their aggregate power is more stable than each individual site. Third, while the variability is high, renewable energy production is predictable for a reasonable 
prediction horizon (i.e., 3--24 hours). 
This offers sufficient time for datacenter operators to identify power changes in advance, such that they can migrate workloads hours ahead for 
tolerating the power variability. 


\name{} develops three major techniques for facilitating the \siteName{} deployment and operations. 
It first quantifies the coefficient of variation of the power production of renewable energy farms based on their historical traces. This enables datacenter operators to identify the sites that have relatively stable power supply and capacity ($\S$\ref{subsec:site-identification}).
After that, \name{} develops a dynamic subgraph identification algorithm that can group individual renewable energy farm sites into subgraphs. Within the same subgraph, 
the power production pattern of each renewable energy farm is complementary. Therefore, the subgraph will deliver stable aggregated power 
production as a whole ($\S$\ref{subsec:subgraph-identification}). 
Based on the identified subgraph, \name{} helps datacenter operators decide the sites for deploying \siteName{s}. \name{} 
also develops a Mixed-Integer Program (MIP) model for enabling optimized placement and migration of virtual machines (VMs) to further minimize the impact of 
the power variability on application performance ($\S$\ref{sec:app-placement}).


We evaluate \name{} with the power traces collected from hundreds of renewable energy farms, and VM traces from modern data centers. We show that, with careful selection of \siteName{} sites and subgraphs, and proper placement of VMs in \siteName{s}, 
\mbox{\name{}} can reduce the total carbon footprint by $46\%$ with low VM migration frequency, in comparison with conventional  datacenter deployment approaches.
Thus, we believe \name{} can complement current data center architectures to meet the compute demand in a more sustainable 
manner. 

As modular data centers have lower construction cost compared to conventional data centers, in combination with the minimal use of batteries,  
\mbox{\name{}} helps datacenter vendors identify the \siteName{s} deployment that incurs low embodied carbon footprint. With its minimized usage of the power grid, it also has minimal 
operational carbon footprint. As \name{} utilizes the aggregated power production of a few stable energy farms, it minimize 
the overprovisioning of power and compute resources, which further reduces both the embodied and operational carbon emission. Overall, we make the following contributions in the paper:


 \begin{itemize}[noitemsep,nolistsep,topsep=0pt,leftmargin=*]
        \item We conduct a characterization study of renewable energy with power traces from real-world energy farms ($\S$\ref{sec:motivation}).
        \item We design a site-pruning technique using historical traces to locate viable renewable sites that fit for 
        rMDCs ($\S$\ref{subsec:site-identification}).
        \item We propose a subgraph identification technique that can identify small complementary subgroups of energy farms based on 
        the prediction of their power variability ($\S$\ref{subsec:subgraph-identification}). 
        \item We develop optimization techniques for smart VM placement and migration to minimize the negative impact of power variability 
        on VM performance and maximize the efficiency of \siteName{s} ($\S$\ref{sec:app-placement}).
        \item We implement \name{} framework to facilitate the deployment of \siteName{s} at scale, and conduct a detailed evaluation 
        using real-world renewable energy and VM traces ($\S$\ref{sec:eval}).
\end{itemize}

\begin{figure}[t]
\centering
  \includegraphics[scale=0.9]{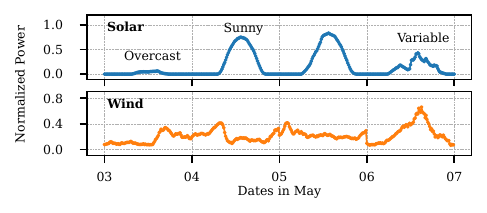}
  \vspace{-3ex}
  \caption{Power variation over time in renewable energy farms.}
  \label{fig:solar-variability}
  \label{fig:wind-variability}
\label{fig:variability}
 \vspace{-3ex}
\end{figure}

\section{Characterization of Renewable Energy}
\label{sec:motivation}


To motivate the design of \name{}, we first characterize the variability of renewable energy sources.
We quantify the variability of two main renewable energy sources, solar and wind, by analyzing two 
representative years-long datasets: (1) EMHIRES dataset~\cite{iratxe2016emhires, gonzalez2017emhires} 
with traces from 555 sites in Europe and (2) ELIA dataset~\cite{elia} from 25 sites in Belgium that 
includes per-site power production traces and forecasts.



As expected, we observe significant variation across space (different farm locations), 
time (times of the day and seasons), and power resources (solar and wind).
Figure~\ref{fig:solar-variability} shows a 4-day sample of solar production, 
normalized to the maximal energy production capacity. 
Solar energy follows a periodic diurnal pattern, but days overcast with heavy clouds can significantly 
reduce the peak production (3.5\% vs. 77\% in the following day), and days with variable cloud patterns 
cause spiky energy production. 
And different energy sources exhibit different patterns. 
The wind energy production for the same duration exhibits sharp peaks and valleys (depending on the 
weather conditions), but rarely drops to zero.



\subsection{Complementary Variability Patterns}
\label{sec:multi-VB}

\begin{figure}[t!]
    \centering
    \begin{subfigure}[t]{0.47\columnwidth}
    \includegraphics[width=\textwidth]{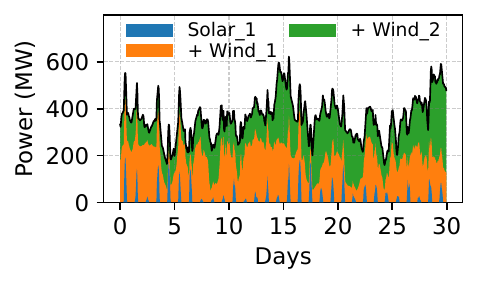}
    \vspace{-5ex}
    \caption{{Complementary power production pattern of three sites.}}
    \label{fig:multi-site-example-a}
    \end{subfigure}%
    \hfill
    \begin{subfigure}[t]{0.47\columnwidth}
        \includegraphics[width=\textwidth]{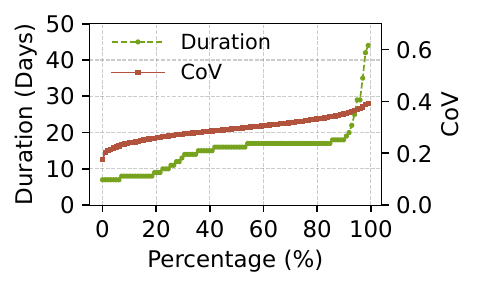}
        \vspace{-5ex}
        \caption{{Distribution of the duration of the complementary patterns.}}
        \label{fig:multi-site-example-b}
    \end{subfigure}
     \vspace{-1ex}
    \caption{Reducing the variability in renewable energy production by aggregating multiple sites.}
    \vspace{-4ex}
    \label{fig:multi-site-example}
\end{figure}

Despite the large variability in a single renewable energy farm, we observe that these variability patterns are often complementary 
among different energy farms across a geographical region. 
This complementary pattern can be generated by using different energy sources (e.g., wind vs. solar), 
and geographical locations with different impacts of micro-climates and weather (e.g., the same solar source 
but in two different locations, one of which is covered by clouds). 

{\mbox{Figure~\ref{fig:multi-site-example-a}}} demonstrates such an example from the EMHIRES dataset.  
To gauge the variability of power produced over time, we use the coefficient of variation (CoV) of renewable power produced at 
different timestamps as the metric. CoV is the standard deviation divided by the mean, so a higher CoV implies more variability. 
In Figure~\ref{fig:multi-site-example-a}, by combining the energy sources from one solar farm (Solar\_1) 
and 
one wind farm (Wind\_1) 
we deliver more stable energy than the single solar farm (5.6$\times$ lower CoV).
Adding another wind farm (Wind\_2) 
further reduces the variability and decreases the value of CoV by an additional 1.5$\times$. Thus, we reach that:  

\vspace{0.5ex}
\noindent\textit{\textbf{\uline{Observation 1:}} Power variability in renewables can be masked by selecting a subgroup
of complementary sites.}
\vspace{0.5ex}

We further analyze how long the complementary sites can remain complementary. 
We define a group of sites as complementary sites if their aggregated power production is stable (CoV$< 0.4$) for at least one week. 
Our results show that the complementary sites remain stable for $14.4$ days on average and up to $44$ days, as shown in Figure~\mbox{\ref{fig:multi-site-example-b}}. 
For the example shown in \mbox{Figure~\ref{fig:multi-site-example-a}}, the three renewable sites  
can maintain a complementary pattern for 30 days. 
Therefore, our analysis suggests that we can re-identify complementary sites periodically but not frequently. 

\vspace{0.5ex}
\noindent\textit{\textbf{\uline{Observation 2:}} 
{The complementary sites could preserve the complementariness for 2 weeks on average and up to 6 weeks.}}


\subsection{Predictability of Renewables}
\label{sec:predictability}

\begin{figure}[t]
    \centering
    \includegraphics[width=0.85\columnwidth, trim={0 0 0cm 0.2cm}]{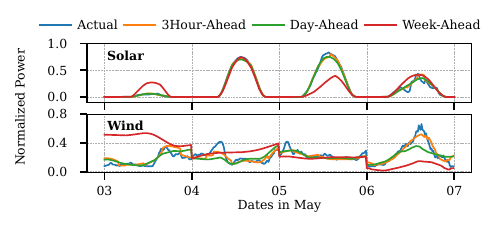}
    \vspace{-3ex}
    \caption{Energy prediction of solar and wind in near (3-hour and day-ahead) and far-away future (week-ahead). }
    \label{fig:energy-prediction}
    \vspace{-3ex}
\end{figure}

The main cause of power variability is the weather condition, which can be predicted accurately a few hours ahead. 
We show the power forecasts provided in the ELIA dataset~\cite{elia} (based on the weather forecasts) in Figure~\ref{fig:energy-prediction}.
The predictions for near-future power production are accurate enough to capture 
important trends. 
The mean absolute percentage error (MAPE) for the next 3-hour predictions is 8.5--9.0\%, for day-ahead predictions is 18--25\%, and for week-ahead predictions is 44\%--75\%. 
This predictability provides insights into how the power resources will change, and when will workload migrations be needed (see the 
detailed discussion in $\S$\ref{sec:app-placement}). As the power changes can be predicted within several hours to one day ahead, we have sufficient time to 
migrate applications to tolerate the power variability further. 

\vspace{0.5ex}
\noindent\textit{\textbf{\uline{Observation 3:}} Renewable energy is predictable for a reasonable prediction horizon (3--24 hours) in the future}. 

\section{Technical Background}
\label{sec:background}
To facilitate our discussion, we now present the essential technical background of modular data centers. 


\subsection{Modular Data Centers}
\label{subsec:mdc}

Modular data centers (MDCs) attract much attention from datacenter vendors, as they have low construction cost and installation time~\cite{hamilton2006mdcarch, vishwanath2009modular, shuja2016sustainable}. An MDC organizes 
the server racks, cooling system, power supply, and batteries in one or more containers, which provides convenience for installation and shipping.  
In Table~\ref{tab:mdc}, we list the configurations and costs of an example MDC FusionModule-2000~\cite{mdc-config}. 
A typical MDC consists of 10 racks (each rack with 15 servers), 3 battery cabinets with a default 15-minute backup time, 
and 3 cooling containers. Among all the components, server racks take a major portion of the embodied carbon (i.e., 
carbon emission of manufacturing, construction, and shipping), installation cost, and building footprint. 
The battery cabinets have less installation cost and building footprint, 
as they are usually used for emergency backups with a small capacity. 



\begin{table}[t]
  \centering
  \caption{A summary of the core components in a modular data center, and their cost and characteristics.}
	\vspace{-1ex}
 \resizebox{.46\textwidth}{!}{
  \footnotesize{
  \begin{tabular}{|c|c|c|c|}
  \hline
  Modular DC               & Server Rack            & Cooling          & Battery         \\ \hline
  \textbf{Power} ($kW$)              & 150        & 35          & -  \\ \hline
  \textbf{Embodied Footprint} ($tCO_{2}e$) & 88.7           & 3.7        & 5.5 \\ \hline
  \textbf{Cost} ($\$$)                  & 450K &     10.9K   & 46.9K        \\ \hline
  \textbf{Footprint} ($m^2$)              & 20.6       & 5.8 & 5.8 \\ \hline
  \textbf{Capacity}                    & 10 racks &    -     & 15 mins    \\ \hline
\end{tabular}
  }}
  \vspace{-4ex}
  \label{tab:mdc}
  \end{table}

\subsection{Renewable Energy-based MDCs}
\label{subsec:rdc}

MDCs are well-suited to collocate with renewable energy farms for their flexibility and low construction costs. Major datacenter providers have recently invested in building \siteName{s} that can be easily deployed alongside renewable energy farms~\cite{microsoftmdc, ibmmdc}. However, the key challenge with 
collocation is the variability of the renewable energy production (see $\S\ref{sec:motivation}$). 
To cope with this, current studies usually leverage batteries and/or power grids. 
We discuss their pros and cons as follows.


\noindent \textbf{{Batteries.} }
Batteries can tolerate the volatility of renewable power supply\mbox{~\cite{carbonexplorer:asplos2023, 
thompson2016optimization, mccluer2008comparing}} by storing the surplus renewable energy and discharging 
it when the renewable power is insufficient. However, deploying batteries in \siteName{s} 
faces two critical challenges. 
First, the current \siteName{} battery capacity is insufficient to tolerate the renewable energy production variation (up to several hours). 
Second, increasing the battery size will increase the cost and building footprint. 
Considering the commercial-scale battery ($\$1250/kWh$\mbox{~\cite{battery_nrel}}) 
with a one-hour backup time (for servers in Table~\mbox{\ref{tab:mdc}}), the hardware and installation cost accounts for $>$30\% of the total \siteName{} cost, and the building footprint reaches $>$40\% of the total footprint. 
Therefore, it is less practical to fully rely on batteries to tolerate
long-lasting renewable power supply fluctuations. 

\noindent \textbf{{Power grid.} }
The power grid serves as a backup energy source for \siteName{s} as it is a more stable power supply. 
Drawing energy from the power grid helps the \siteName{} keep servers running when the renewable power supply drops. 
However, the grid mainly supplies power from non-renewable energy sources today, which have significantly larger carbon intensity\mbox{~\cite{gupta2021chasing}}. 
Therefore, extensive power grid use will increase the operational carbon footprint of the \siteName{}.

\section{Design and Implementation}
\label{sec:design}


The goal of \name{} is to best utilize renewable energy when collocating \mdc{s} with 
renewable energy farms, while minimizing the negative impact on application performance. 
\name{} maximizes the use of renewable energy for \siteName{s} and minimizes the use of power grid and batteries to reduce carbon emissions when renewable energy is 
insufficient. 

\subsection{\name{} Overview}
\label{subsec:overview}

\begin{figure}[t]
    \centering
    \includegraphics[scale=0.435]{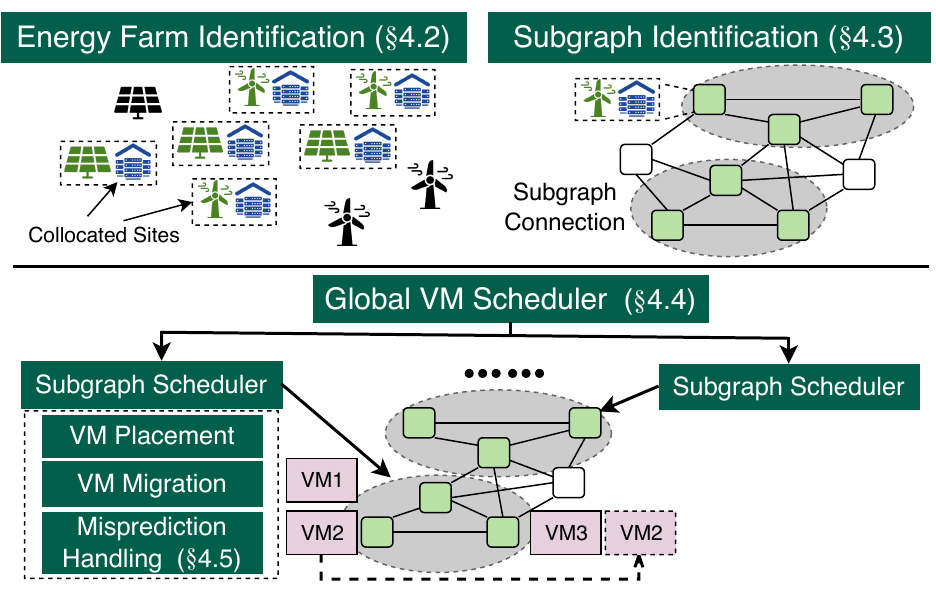}
    \vspace{-4ex}
	 \caption{System overview of \name{}.
     \vspace{-4ex}
    }
    \label{fig:overview}
\end{figure}

\name{} has four 
core components as shown in Figure \mbox{\ref{fig:overview}}.

\noindent \textbf{Identification of renewable energy farms:} To decide where to 
deploy \siteName{s}, \name{} uses the stability of power supply as the key metric. 
By collocating \siteName{s} with stable renewable energy farms, it ensures a reliable power supply while reducing embodied carbon by overprovisioning fewer servers (\S\ref{subsec:site-identification}).

\noindent \textbf{Subgraph identification of complementary sites:} Our study reveals 
that a small group of renewable energy farms with complementary patterns can produce stable aggregated 
power supply ($\S$\ref{sec:multi-VB}). \name{} identifies a set of subgraphs from the selected \siteName{s}, where each of them has complementary renewable energy soures. 
With the subgraph candidates, we select ones with the stablest power supply ($\S$\ref{subsec:subgraph-identification}). 

\noindent \textbf{VM placement and migration:}
Since each individual \siteName{} of a stable subgraph may still face power variability,
\name{} performs optimzied VM placement and migration within the subgraph. 
Utilizing the predictability of renewable energy production (see $\S$\ref{sec:predictability}) 
and VM characteristics (e.g., lifetime)~\cite{protean},
\name{} can place VMs on more stable \siteName{} and perform VM migrations from \siteName{s} with insufficient power 
to those with excessive power. This minimizes the power grid usage, and thus reduces the operational carbon footprint
($\S$\ref{sec:app-placement}).

\noindent \textbf{Misprediction handler:} Predictions of the future power supply and VM lifetime 
may incur misprediction errors. 
\name{} gracefully handles the mispredictions with minimal overhead. We will discuss the details in $\S$\ref{subsec:misprediction}.

\subsection{Identification of Energy Farms for \siteName{s}}
\label{subsec:site-identification}
\name{} first decides which renewable energy farm to collocate with each \siteName{}.
There exists a large number of renewable energy farms, but not all of them are suitable to 
deploy \siteName{s}. This is because different farms have different levels of power variability. 
\name{} prefers renewable energy farms with stable power supply for two major reasons. 

First, collocating \siteName{s} with stable farms guarantees higher resource availability, which reduces the power grid usage and incurs less VM outages. 
Second, the collocation scheme helps reduce the embodied carbon footprint and the demand for batteries. This is because \siteName{} with a higher power supply fluctuation 
has to overprovision additional server capacity to utilize the peak renewable energy production. \siteName{s} with more stable power supply require 
less server capacity and battery capacity, therefore mitigating the embodied carbon footprint and construction cost.  

\begin{figure}[t]
    \centering
    \includegraphics[width=0.45\textwidth, ,trim={0 0 0 0.3cm}]{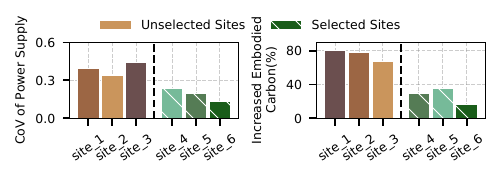}
    \vspace{-3ex}
	 \caption{Improved power stability and reduced embodied carbon footprint with selected energy farms (lower is better). 
     }
    \vspace{-3ex}
    \label{fig:site_pruning}
\end{figure}

\name{} uses the coefficient of variation (CoV) of the power production 
as the key metric to rank the farms.
Figure~\ref{fig:site_pruning} (left) shows the CoV of the three selected energy farms is much less (51\%) than that of 
the selected energy farms. In comparison with \siteName{s} supplied with stable power sources (e.g., power grid), Figure~\ref{fig:site_pruning} (right) shows
the additional embodied carbon cost of building \siteName{s} of these selected farm sites is much less than that of unselected farms (see the detailed procedure of calculating carbon emissions in $\S$\ref{sec:setup}). 
This is because these \siteName{s} need less servers for the overprovisioning 
for tolerating the power variability of collocated energy farms. 
Although \name{} causes embodied carbon 
footprint for constructing \siteName{s}, it significantly reduces the total carbon emissions for \siteName{s} by reducing 
the operational carbon footprint (see $\S$\ref{sec:eval}). 


%

\subsection{Complementary Subgraph Identification}
\label{subsec:subgraph-identification}


In $\S$\ref{subsec:site-identification}, we identify a set of renewable energy farms with stable power supply to collocate with \siteName{s}.
However, three major challenges remain. 
First, managing these \siteName{s} across multiple geographical regions will be 
complicated. It brings challenges to datacenter operators, especially considering the diverse power production patterns at different farms. 
Second, the selected farms individually cannot constantly generate stable power production over time.
Third, as cloud providers place their VM workloads across 
all the \siteName{s}, the decision space could be extremely large. An ill-judged decision could cause the inefficiency of VMs, 
and even unexpected VM outages due to the unstable power supply.

\noindent\textbf{\underline{Key idea:}}
To address those challenges, we identify disjoint subgraphs from these \siteName{s}. 
Each subgraph has a subset of renewable energy farms. 
To build a subgraph, we use three metrics: 
(1) \textit{the CoV of the aggregated power production}; 
(2) \textit{the minimum power production of the farms};
and (3) \textit{the distance between sites}. 
The first and second metrics are based on the historical power production of each farm. They guarantee that we select the subgraphs with 
sufficient and stable aggregated power supply. The third metric 
ensures the selected subgraphs have low communication overheads. It is configurable, and we set the upper bound as 500 miles by default. 
We enumerate all possible subgraphs and rank them 
using the three metrics.
We iterate through the set of subgraphs in the rank order and select the subgraph that has no intersection with previously selected subgraph. 
After that, we have a set of subgraphs that have complementary patterns.

%

\begin{figure}[t]
    \centering
    \includegraphics[width=.45\textwidth,trim={0 0.3cm 0 0.3cm}]{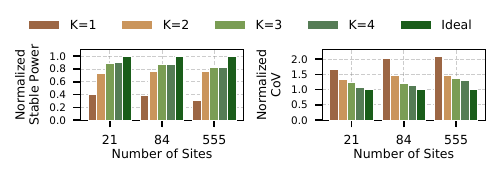}
    \vspace{-0.5ex}
    \caption{The impact of different subgraph size $K$. It shows the normalized stable aggregated power and its average CoV.}
    \label{fig:design:sub-graph-size}
    \vspace{-3ex}
\end{figure}

\noindent\textbf{\underline{Key study results:}}
Exhaustively enumerating all the subgraph candidates can lead to non-trivial overhead, since the number of subgraph candidates grows combinatorially as we increase 
the number of farm sites and the size of subgraphs. 
Ideally, we wish to minimize the size of subgraphs, while retaining the 
complementary benefits,  
which can simplify the power management and task scheduling across data centers.   

\noindent\textbf{Small subgraphs are sufficient for \siteName{} deployment:}
According to our study of the 555 farms, 
we find that small subgraphs with three sites have complementary power patterns, and can provide a sustained power capacity. 
We show the aggregated power and average 
CoV of the power with different subgraph sizes in Figure~\ref{fig:design:sub-graph-size}. In the best case (Ideal in Figure~\ref{fig:design:sub-graph-size}), 
all farm sites are grouped into a single subgraph and it has the highest stable aggregated power and the lowest variability (CoV). 
As we vary the size of subgraphs in Figure~\ref{fig:design:sub-graph-size}, the stable aggregated power of the subgraphs with three sites 
can reach the 80-90\% of the ideal stable power, and its CoV is also comparable to the ideal case. 
As we further increase the subgraph size, we will enlarge the exploration space without significant benefits. 
Therefore, we set the subgraph size to be three by default for the studies in \name{}.

\noindent\textbf{Subgraph identification is a dynamic problem:}
As the power patterns of a renewable site may change over time. 
Our study shows that subgraphs usually retain their 
complementariness for two to six weeks. 
Therefore, \name{} can re-identify  subgraphs that have complementary power patterns every two weeks for more stable aggregated power supply. 

%

As each subgraph has stable aggregated power supplies from renewables, 
datacenter vendors can deploy \siteName{s} close to renewable energy sites in these subgraphs. Note that datacenter vendors have the choice to select one or more 
subgraphs for \siteName{} deployment. They can deploy \siteName{s} at the farm sites in the selected subgraphs. 
After the \siteName{} deployment, we do not need to move the deployed \siteName{s} (even upon subgraph re-identification). 
Instead, \name{} suggests migrating VMs across \siteName{s} for lower cost (see the details in $\S$\ref{sec:app-placement}).

\subsection{VM Placement and Migration}
\label{sec:app-placement}
After identifying the subgraphs, we now place VMs across these subgraphs and their \siteName{s}. 
A simple approach is heuristic-based VM placement, such as greedily placing VMs onto the \siteName{s} in the order of their resource availability. 
However, heuristic-based approaches are often suboptimal, especially when the power supply changes over time, and different VMs have different priorities 
and demands for computing resources. Therefore, it is difficult for heuristic-based approaches to identify the best \siteName{} for each VM.

\name{} formulates the VM placement and migration problem into an optimization problem using the Mixed-Integer Program (MIP) model~\cite{mip}. \name{} employs the MIP model for three reasons. 
First, the MIP model can make optimized decisions for a large set of VMs and \siteName{s} with a global perspective. 
Second, it can take VM properties (e.g., regular or evictable) 
into consideration to match each VM with its best-fit \siteName{}. Third, the MIP model can effectively handle the dynamic power supply patterns of \siteName{s} 
by leveraging the predictability of renewable power supply to find optimized solutions in advance.
We present the VM placement and migration policies using a hierarchical approach as follows.
\begin{table}[t]
    \centering
    \caption{Input constants of our MIP model.}
	\vspace{-1ex}
    \resizebox{.46\textwidth}{!}{
    \footnotesize{
    \begin{tabularx}{\linewidth}{|c|X|}
        \hline
        Symbol & Interpretation \\
        \hline
        ${\mathbb{M}, \mathbb{D}, \mathbb{N}, \mathbb{K}_n, \mathbb{T}}$ & A set of $M$ regular VMs, $D$ delay-insensitive VMs, N subgraphs, $K$ \siteName{s} for subgraph $n$, time interval T \\\hline
        $Power_{mt}$ & The power draw of VM $m$ at time $t$ \\\hline
        $Mem_{m}$ & The memory size of VM $m$ \\\hline
        $Lifetime_{m}$ & The predicted lifetime of VM $m$ \\\hline
        $RS_{nkt}$ & Renewable power supplied to the $k$th \siteName{s} of subgraph $n$ at time $t$ \\\hline
	    $Power_{Migr}$ & Power consumption of migration (per GB VM states) \\\hline
        $CI_{NR}, CI_{R}$ & Carbon Intensity of non-renewable energy and renewable energy~\cite{carbonexplorer:asplos2023}\\\hline
    \end{tabularx}
    }
    }
    \label{tab:constants}
     \vspace{-4ex}
\end{table}

\noindent \textbf{{VM placement among subgraphs:} }
\name{} first relies on heuristic-based approaches for VM placement across subgraphs. This is because the aggregated power supply of each subgraph 
is relatively stable. We empirically find that the heuristic-based approach can handle such scenario well. Specifically, \name{} employs 
the best-fit placement algorithm. We maintain a list of subgraphs ordered by the amount of excess power. 
When a VM arrives, we place it on the subgraph with the highest amount of resource available. If the subgraph has no excess power supply, 
we remove it from the list.

\noindent \textbf{{VM placement among \siteName{s} in a subgraph:}}
After identifying the suitable subgraph, \name{} decides where to place the VM among \siteName{s} within each subgraph. 
Since the renewable power supply for each individual \siteName{} is less stable than the aggregated power supply of each subgraph, 
the VM placement problem across \siteName{s} is more dynamic. 
Therefore, we do not use a heuristic-based approach, but instead formulate it into an optimization problem using the MIP model. 
We define the input constants and variables of MIP model in Table~\ref{tab:constants} and Table~\ref{tab:symbols}, respectively, and explain them below.

\begin{footnotesize}
\vspace{-3ex}
\begin{align}
\label{eq:subgraph_placement}
\text{O1 Carbon:} &\, \textbf{min} \sum_{n \in \mathbb{N}, k \in \mathbb{K}_n, t \in \mathbb{T}} NR_{nkt} \cdot CI_{NR} + RU_{nkt} \cdot CI_{R} \\
\text{O2 Uptime:}
&\, \textbf{max} \sum_{m \in \mathbb{D}}{\frac{Lifetime_m}{Lifetime_m + Downtime_m}}, \quad \text{\bf s.t.}  \nonumber \\ 
\text{C1 Power:} & \, \forall n \in \mathbb{N}, k \in \mathbb{K}_n, t \in \mathbb{T}: \nonumber \\
& \, RU_{nkt} \leq RS_{nkt} \nonumber \\ \nonumber
& \, Consum_{nkt} = NR_{nkt} + RU_{nkt} = \sum_{m \in \mathbb{M\cup D}} X_{mnt} \cdot Power_{mt} \nonumber \nonumber \\ \nonumber
\end{align}
\vspace{-6ex}
\end{footnotesize}
\renewcommand{\lstlistingname}{Formula}
\begingroup\vspace*{-\baselineskip}
\vspace{-1ex}
\captionof{lstlisting}{The key objectives and constraint of MIP model.}
\vspace{-1ex}
\vspace*{\baselineskip}\endgroup

\noindent
\uline{MIP input:} 
The MIP model takes the following inputs: 
(1) VM resource configuration, including its memory size ($Mem_m$), its VM category (regular or evictable\footnote[1]{Cloud platforms usually offer evictable VMs that run at a much lower price and priority than regular VMs. They can be evicted if needed~\cite{awsspot, azurespot}.}), 
and its power consumption ($Power_{mt}$), which can be estimated by the number of vCPUs and its utilization ~\cite{kansal2010virtual};
(2) Estimated VM lifetime ($Lifetime_m$), based on the insights from prior studies~\cite{protean, resourcecentral, barbalho2023virtual} 
that VM lifetime can be estimated with high accuracy using VM properties (e.g., VM type and hosted application type);
(3) Current and predicted power supply of \siteName{s} ($RS_{nkt}$), with the high predictability of renewable power supply (see $\S$\mbox{\ref{sec:predictability}}). \name{} has the strong robustness of handling 
mispredictions, as discussed in $\S$\ref{subsec:misprediction} and $\S$\ref{sec:eval:misprediction}.

\noindent
\uline{MIP output:} 
For VM placement within subgraphs, the MIP model denotes placement decisions as $X_{mnkt}$ (see Table~\ref{tab:symbols}). 
It sets $X_{mnkt}=1$, if VM $m$ is powered on at $k$th \siteName{} in subgraph $n$ at time $t$. 
The MIP model can also decide the non-renewable energy usage from the power grid. We use $NR_{nkt}$ to represent the 
amount of non-renewable power supply to the $k$th \siteName{} in subgraph $n$ at time $t$.

\noindent
\uline{MIP objectives:} \name{} aims to minimize the total carbon footprint and maximize the VM uptime.
To achieve these goals, we define two objectives in our MIP model. The first objective (O1 Carbon in Formula~\ref{eq:subgraph_placement}) is 
defined to minimize the non-renewable energy usage of \siteName{s} over the course of time, since it is the primary cause of 
the operational carbon footprint. The second objective maximizes the uptime percentage for evictable VMs (O2 Uptime). 
As for regular VMs, \name{} guarantees that they will never be actively powered off due to the insufficient renewable power supply.

\noindent
\uline{MIP variables and constraints:} We list the key constraint of our MIP model in Formula \ref{eq:subgraph_placement}. 
It enforces that the total power consumption of all running VMs ($Consum_{nkt}$) on each \siteName{} cannot exceed the total power supply (C1 Power).

\begin{table}[t]
    \centering
    \caption{Variables used in our MIP model.}
	\vspace{-1ex}
 \resizebox{.46\textwidth}{!}{
    \footnotesize{
    \begin{tabularx}{\linewidth}{|c|c|X|}
        \hline
        Symbol & Domain & Interpretation \\
        \hline
        $X_{mnkt}$ & $\{0, 1\}$ & Whether VM $m$ is powered up in the $k$th \siteName{} of the subgraph $n$ at time $t$ \\\hline
        $M_{mnk_1k_2t}$ & $\{0, 1\}$ & Whether VM $m$ is migrated from the \siteName{} $k_1$ to $k_2$ of the subgraph $n$ at time $t$\\\hline
        $NR_{nkt}$ & $\mathbb{R}_{\ge 0}$ & Non-renewable power used by the $k$th \siteName{s} of the subgraph $n$ at time $t$\\\hline
        $RU_{nkt}$ & $\mathbb{R}_{\ge 0}$ & Renewable power used by the $k$th \siteName{s} of the subgraph $n$ at time $t$\\\hline
        $Consum_{nkt}$ & $\mathbb{R}_{\ge 0}$ & Power consumption in the $k$th \siteName{s} of the subgraph $n$ at time $t$\\\hline
        $Downtime_{m}$ & $\mathbb{R}_{\ge 0}$ & The actual downtime of VM $m$\\ \hline
        
        
    \end{tabularx}
    }}
    \vspace{-1ex}
    \label{tab:symbols}
\end{table}

\noindent \textbf{VM migration across \siteName{s} in a subgraph:} If an \siteName{} experiences a decrease of renewable power supply, 
\name{} can migrate some of its VMs to another \siteName{} within the same subgraph that has excessive power supply. Therefore, 
we integrate a power model to quantify the overhead of VM migration, where its power consumption is proportional to the size of VM 
states migrated~\cite{vm_migration_hpdc11}. Larger VMs incur higher overhead and are more sensitive to migrations.

\noindent
\uline{Additional MIP output:} 
VM migration is represented by $M_{mnk_1k_2t}$ (Table~\ref{tab:symbols}) in our MIP model. $M_{mnk_1k_2t}$ denotes whether a VM $m$ should be migrated 
from $k_1$th \siteName{} to $k_2$th \siteName{} of subgraph $n$ at time $t$.
Using the same objectives (O1 and O2), the MIP model will identify an optimized VM migration plan for each \siteName{}.

\noindent
\uline{Additional MIP variables and constraints:} 
The MIP model uses $M_{mnk_1k_2t}$ and input constant $Power_{Migr}$ to estimate the migration overhead of a VM $m$. For each VM, its migration overhead counts toward the total power consumption of both migration source and target \siteName{} (C1' Power of Formula~\ref{eq:formula_2}). Following the prior constraint (C1 Power), the total consumption cannot exceed the power supply.

\begin{footnotesize}
\vspace{-1em}
\begin{align}
\text{C1' Power:} & \, \forall n \in \mathbb{N}, k \in \mathbb{K}_n, t \in \mathbb{T}: \nonumber \\
&  Consum_{nkt} = \! \!\!\!\! \sum_{m \in \mathbb{M\cup D}} [X_{mnkt} \cdot Power_{mt} + Power_{Migr} \cdot \nonumber \\
& \quad \quad \quad \quad Mem_m \cdot (\sum_{k_1 \in \mathbb{K}_n} M_{mnk_1kt}  + \sum_{k_2 \in \mathbb{K}_n} M_{mnkk_1t}) ] \label{eq:formula_2} \\ \nonumber
\end{align}
\vspace{-6ex}
\end{footnotesize}
\begingroup\vspace*{-\baselineskip}
\vspace{-1ex}
\captionof{lstlisting}{The additional constraint of the MIP model.}
\vspace{-3ex}
\vspace*{\baselineskip}\endgroup

\subsection{Misprediction Handling}
\label{subsec:misprediction}

Our MIP model relies on the prediction of renewable power production and VM lifetimes, which may be slightly inaccurate.
For instance, a 3-hour-ahead prediction has $9\%$ error on average.
In \mbox{\name{}}, minor prediction errors are tolerable. While the actual power supply of an \siteName{} may be lower than predicted, 
it may still exceed the actual power consumption. And large mispredictions may cause extra power grid usage or evictable VM outages.
Therefore, we design a misprediction handling mechanism to minimize their impact.

 \begin{table}[t]
    \centering
    \scriptsize
        \caption{Misprediction handling for different scenarios.}
	\vspace{0ex}
    \label{tab:exceptions}
    \vspace{-1ex}
        \begin{tabular}{|p{5pt}<{\centering}|p{40pt}<{\centering}|p{56pt}<{\centering}|p{100pt}<{\centering}|}
            \hline
            \textbf{ID} & \textbf{VM Lifetime} & \textbf{Renewable Power} &  \textbf{Possible Exceptions} \\
            \hline
                \circleb{1} & Over-predict & Under-predict & N/A\\
            \cline{1-4}
                \circleb{2} & Over-predict & Over-predict & Power Deficiency\\
            \cline{1-4}
                {\circleb{3}} & Under-predict & Under-predict & Power Deficiency, Extra VM Lifetime\\
            \cline{1-4}
                \circleb{4} & Under-predict & Over-predict & Power Deficiency, Extra VM Lifetime \\
            
            \hline
        \end{tabular}
        \vspace{-5ex}
\end{table}

\begin{figure}[t]
  \centering
  \includegraphics[width=0.43\textwidth, trim={0 0 0cm 0}]{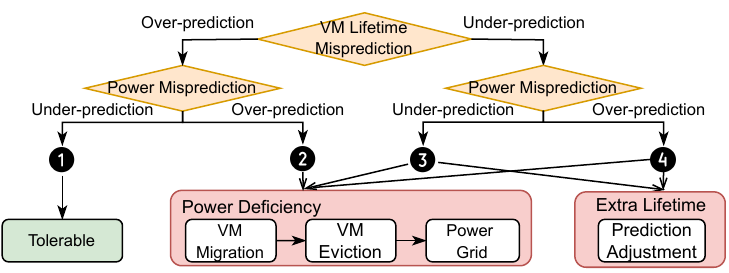}
	 \vspace{-1ex}
	 \caption{The misprediction handling mechanisms of \name{}.}
  \label{fig:design:workflow}
	 \vspace{-3ex}
\end{figure}

We categorize the misprediction into two types: under-prediction and over-prediction, where an under-predicted value is smaller than its actual value, and vice versa. 
For example, an under-predicted power production means more actual power is produced than predicted, and an over-predicted VM lifetime means the VM's actual lifetime is shorter than predicted.
We present different misprediction scenarios in Table~\ref{tab:exceptions}, and the workflow of misprediction handling in Figure~\ref{fig:design:workflow}.

In \circleb{1}, the actual VM lifetime is shorter than predicted,  and the actual power supply is larger than predicted. 
Based on the predicted values, the MIP model may place fewer VMs on the \siteName{} than it can actually support, and these VMs may run shorter 
than its predicted lifetime. In this case, the power consumption never exceeds the power supply. The \siteName{} will have excessive power, and 
it can serve as the migration target of other \siteName{s}. Thus, no specific handling needed for it. 

In \circleb{2}, 
both the VM lifetime and the renewable power supply are over-predicted.
Since the power supply is over-predicted, our MIP model may place more VMs on the \siteName{} than that it can support, 
resulting in an power deficiency at the \siteName{} (i.e., the power supply at an \siteName{} cannot meet its total power consumption). 
To handle this, we employ three techniques: \textit{VM migration}, \textit{evictable VM shutdown}, and \textit{power transmission from the grid}. 
First, we prioritize VM migration since it introduces minimal VM downtime and carbon overhead.
To perform migration, we select another \siteName{} which has the most excessive power supply in the same subgraph as the migration target. 
If such an \siteName{} exists, we will migrate VMs to this target \siteName{} until it cannot host more VMs, or the power consumption 
of the source \siteName{} falls below the power supply.
We prioritize the VMs with less amount of VM states to migrate, as they incur lower migration overheads.
If the power consumption still exceeds the power supply, we will power off evictable VMs until their VM uptime percentage 
is below a predefined threshold (empirically set to 90\%). 
Finally, we will use the power grid if still necessary.

In \circleb{3}, the actual VM lifetime is longer than its prediction. The MIP model might place VMs onto the \siteName{} that does not have sufficient power supply for its entire lifetime. 
This problem can be handled with similar mechanisms in \circleb{2}. Since the MIP model receives the under-predicted VM lifetime as its input, 
it cannot make placement and migration decisions for the rest of their actual lifetimes. 
To handle this issue, \name{} adds an offset to the VM lifetime prediction when we detect that it is under-predicted. This offset is adjusted using the average difference between the predicted and the actual lifetime of completed VMs.

In \circleb{4}, it may incur similar problems caused by over-predicted power supply in \circleb{2} and under-predicted VM lifetime in \circleb{3}. We handle them in the same way as discussed above.

\subsection{\mbox{\name{}} Implementation}
\label{subsec:implementation}

We develop \name{} framework by enabling the replay of the power variations for each \siteName{} with real-world power traces1 from renewable energy farms, and the replay of the VM execution and migration across \siteName{s} with real-world VM traces from data centers~\cite{azure_vm_traces}. 
It takes renewable power traces and VM traces as its inputs and replays the power supply in each \siteName{} and the VM events (i.e., VM deployment, migration, or shutdown). We measured the VM migration overheads for different types of VMs in real servers ($\S$\ref{sec:app-placement}). 
At runtime, \name{} updates the power production at each \siteName{}, 
triggers VM placements, and decides VM migration. 
We use Gurobi~\cite{gurobi} to implement our MIP model for VM placement and migration, the MIP model uses a 3-hour-ahead prediction for its inputs such as the future power supply.
The migration decisions of each subgraph are executed concurrently. 

\section{\name{} Evaluation}
\label{sec:eval}

Our evaluation shows that: (1) \name{} reduces the total carbon footprint by $46\%$ ($\S$\ref{sec:eval:carbon}) 
with reduced monetary cost ($\S$\ref{sec:eval:cost}),  
and minimizes its impact on VMs, compared with baseline \siteName{} placement policies ($\S$\ref{sec:eval:downtime}); 
(2) It remains performant with large-scale \siteName{s} deployment, different battery capacities, and various VM workloads ($\S$\ref{sec:eval:sensitivity}); 
(3) It minimizes the use of power grid for reduced carbon emissions ($\S$\ref{sec:eval:breakdown}); 
and (4) it shows strong robustness to the mispredictions of renewable power production and VM lifetime ($\S$\ref{sec:eval:misprediction}).    

\begin{table}[t]
    \fontsize{8pt}{8pt}\selectfont
    \centering
    \vspace{3ex}
    \caption{Key parameters used in the evaluation of \name{}.}
    \vspace{-1ex}
    \resizebox{0.48\textwidth}{!}{%
    \begin{tabular}{|c|c|c|c|}\hline
        Category & Component & Value & Lifetime \\\hline
        Cost~\cite{barroso2019datacenter, battery_nrel, josep:icdcs2014} & Server  &  \$3,000 per server & 4 years  \\
        & Battery & \$1,250 per kWh & 10 years \\
        & Power Trans. & \$300K per km & 20 years \\
        & \multirow{2}{*}{\begin{tabular}[c]{@{}c@{}}Construction\\(Cooling, electricity, etc.) \end{tabular}} & \multirow{2}{*}{\begin{tabular}[c]{@{}c@{}}\$10 per watt \end{tabular}} & \multirow{2}{*}{\begin{tabular}[c]{@{}c@{}}20 years \end{tabular}}\\
        & &  & \\ 
        \hline
        \multirow{2}{*}{\begin{tabular}[c]{@{}c@{}}Carbon\\Intensity~\cite{gupta2021chasing} \end{tabular}} & Solar & 41 gCO2eq per kWh & - \\
        & Wind  &  11 gCO2eq per kWh & -  \\
        & Brown & 700 gCO2eq per kWh & - \\\hline
        \multirow{2}{*}{\begin{tabular}[c]{@{}c@{}}Embodied\\Footprint~\cite{hpecarbonfootprint, em2019lithium, rodriguez2019embodied} \end{tabular}} & Server & 591 kgCO2eq per server & 4 years \\
        & Battery & 146 kgCO2eq per kWh & 10 years \\
        & Cooling Facility & 50	kgCO2eq per $m^2$ & 20 years \\\hline
    \end{tabular}
    }
    \label{tab:parameter}
    \vspace{-5ex}
\end{table}

\subsection{Experimental Setup}
\label{sec:setup}




We use the EMHIRES dataset \cite{iratxe2016emhires, gonzalez2017emhires} to obtain the power production data of renewable farms.
Similar to the prior work~\cite{greenhadoop, goiri2011greenslot}, 
we take a constant portion of the power generated from each renewable energy farm, such that its maximum power capacity equals 
to the maximum power consumption of \siteName{} servers. 
We use the VM workloads from the Azure Cloud Dataset\mbox{~\cite{azure_vm_traces}}. The dataset has various VM properties, 
including the VM configurations (e.g., memory size), VM lifetime, and its category (regular or evictable). 
Their arrival rates maintain a high power utilization ($90\%$) over datacenter servers~\cite{greenhadoop, goiri2011greenslot}. 
By default, the VM dataset has $10\%$ evictable VMs.
We also vary the proportion of evictable VMs in our sensitivity analysis (\S\ref{sec:eval:sensitivity}).

\noindent
\textbf{\siteName{} setup:} 
We first obtain the top six \siteName{s} based on the CoV of the power supply and group them into two stablest subgraphs (see \S\ref{subsec:site-identification} and \S\ref{subsec:subgraph-identification}). Each \siteName{} follows the configurations shown in Table~\ref{tab:mdc}.  
We further evaluate \name{} with an increasing number of \siteName{s} in \mbox{\S\ref{sec:eval:sensitivity}}.

\noindent
\textbf{Carbon footprint model:}
We quantify the total carbon footprint with three categories: amortized embodied carbon, operational carbon from the power grid, and operational carbon 
from renewable energy. The embodied carbon footprint is measured with servers, batteries, and cooling facility, and it is amortized over their lifetime. 
We list their details in Table~\ref{tab:parameter}.
The operational carbon footprint is measured by the product of total energy consumption (from wind, solar, or power grid) and the carbon intensity (see Table~\ref{tab:parameter}). 

\noindent
\textbf{Baseline policies:}
We compare \name{} against baseline \siteName{} placement policies. 
They differ in the way that \siteName{s} are placed with multiple energy sources (i.e., renewable energy, battery, and power grid). 
For all baseline policies, \siteName{s} prioritize to use the renewable energy and use the power grid as a backup. 
Note that we do not compare non-renewable-based data centers, as they incur a high operational carbon footprint. 
We summarize the baseline policies as follows:

\begin{figure*}[t]
  \centering
  \includegraphics[scale=0.9]{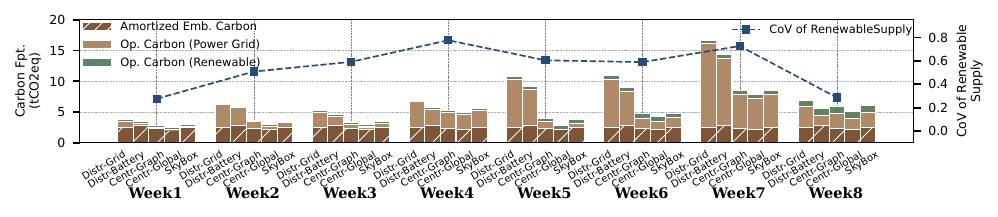}
	 \vspace{-3ex}
	 \caption{The breakdown of carbon footprint with different \siteName{} placements in eight weeks (Emb.: Embodied; Op.: Operational).}
   \vspace{-5ex}
  \label{fig:eval:overall_carbon}
\end{figure*}

\begin{figure}[t]
  \centering
  \includegraphics[scale=0.95]{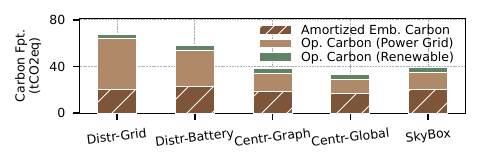}
	 \vspace{-4ex}
	 \caption{Comparison of the cumulative carbon footprint with different \siteName{} placement policies over eight weeks.}
  \label{fig:eval:total_carbon}
	 \vspace{-3.5ex}
\end{figure}

\begin{itemize}[topsep=0pt, noitemsep, leftmargin=*]
    \item \textbf{Centr-Global} deploys a centralized \siteName{} at the geometric center of all the selected renewable energy farms and connects it to all the farms 
	    with power transmission lines. 
    \item \textbf{Centr-Graph} is similar to Centr-Global, but it deploys a centralized \siteName{} per subgraph. They are ideal policies as they can aggregate all the renewable energy into a \siteName{}.
    \item \textbf{Distr-Grid} co-locates the \siteName{} with each renewable energy farm. Each \siteName{} handles its VM workloads independently without migrating VMs. 
    \item \textbf{Distr-Battery} adds extra batteries based on Distr-Grid. The battery charges excessive renewable power and discharges when renewable power is insufficient. We set the extra battery capacity of each \siteName{} such that it can sustain one hour of server operation~\cite{kontorinis2012managing}. We further vary its battery capacity in the sensitivity analysis ($\S$\ref{sec:eval:sensitivity}).
		 
		
\end{itemize}

\name{} is similar to Distr-Grid but live-migrates VMs within each subgraph from \siteName{s} having insufficient power to those with excessive power supply using the MIP model. 
We also o compare \name{} with its variants to show the benefit of each of its component: \textbf{\name{}-NoSI} uses a random site selection policy; 
		\textbf{\name{}-NoSG} uses random subgraph identification; and \textbf{\name{}-BestEffort} uses the best-effort VM placement policy instead of the MIP model. \name{}-BestEffort prioritizes migrating VMs to the \siteName{} with the most excessive power production, suspends evictable VMs if needed, and uses the power grid as the last choice.



  

\subsection{Carbon Footprint Reduction}
\label{sec:eval:carbon}
We first evaluate \name{} in reducing the carbon footprint. 
We show the cumulative carbon footprint of different \siteName{} placement policies over eight weeks in Figure~\ref{fig:eval:total_carbon}.
Compared to Distr-Grid and Distr-Battery, \name{} produces $46\%$ and $39\%$ less total carbon footprint, respectively. 
This is because (1) \name{} groups \siteName{s} into subgraphs, which share a more stable aggregated power supply than each individual \siteName{}; 
and (2) it performs effective VM migration within each subgraph with the predictions of the power supply. 

\begin{figure}[t]
  \centering
  \includegraphics[scale=0.95]{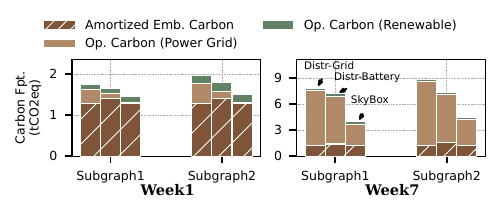}
	 \vspace{-3ex}
	 \caption{The breakdown of carbon footprint of each subgraph.}
  \label{fig:eval:subgraph_carbon_breakdown}
	 \vspace{-5ex}
\end{figure}

We also compare \name{} with the two centralized \siteName{} placement policies: Centr-Global and Centr-Graph.
Since Centr-Global can aggregate all renewable energy production into a centralized data center, it has the most stable power supply.
Compared to this ideal placement policy, \name{} has slightly higher operational carbon, as it incurs extra migration overhead. 
\name{} also produces slightly more embodied carbon, as it needs to provision more servers to make use of the peak power supply. 
Compared to Centr-Graph, \name{} has almost the same ($0.1\%$ more) total carbon footprint. 
This is because Centr-Graph statically connects \siteName{s} with the renewable power farms in each subgraph. 
While these two baseline policies have a similar carbon footprint as that of \name{}, their monetary costs are several times 
higher than that of \name{} (see $\S$\ref{sec:eval:cost}), making them less attractive. 

\name{} achieves more carbon footprint reduction, when the power production is unstable. 
As shown in Figure~\ref{fig:eval:overall_carbon}, \name{} obtains more benefits during the fifth to the seventh week, 
when the CoV of energy production is high (i.e., 51\%-64\% reduction compared to Distr-Grid). 
This is because the baselines require more power grid usage when renewable production is unstable, 
causing a high operational carbon footprint. In contrast, \name{} can perform optimized VM placement 
and migration, which significantly reduces the operational carbon footprint caused by the use of the power 
grid, as shown in the carbon breakdown in Figure~\ref{fig:eval:subgraph_carbon_breakdown}. 

{
\setlength{\tabcolsep}{0.3em}
\begin{table}[t]
    \fontsize{7pt}{7pt}\selectfont
    \centering
    \vspace{3ex}
    \caption{Breakdown of the amortized monetary cost of different data center placement policies over one year. 
    \vspace{-1ex}
    }
    \begin{tabular}{|c|ccccc|}\hline
        Category & Centr-Global & Centr-Graph & Distr-Grid & Distr-Battery & \name{}\\\hline\hline
        Servers  & \$562K & \$614K & \$675K    & \$675K    & \$675K\\\hline
        Battery & \$23K & \$25K & \$28K     & \$141K    & \$28K \\\hline
        Power Trans. & \$62M & \$61M & \$45K & \$45K         & \$45K\\\hline
        Construction & \$375K & \$410K & \$450K    & \$450K     & \$450K \\\hline
        Total   & $\sim$\$63.0M & $\sim$\$62.0M & $\sim$\$1.2M  & $\sim$\$1.3M & $\sim$\$1.2M\\\hline 
    \end{tabular}
    \label{tab:cost}
    \vspace{-4ex}
\end{table}
}

\subsection{Monetary Cost Reduction}
\label{sec:eval:cost}

We evaluate the monetary cost of \name{} amortized over one year, and compare it with the baselines. 
With the monetary cost breakdown and the expected lifetime for each \siteName{} component in Table~\ref{tab:parameter}, 
we show our study results in Table~\ref{tab:cost}. 

\name{} and Distr-Grid have the minimum monetary cost among all the \siteName{} placement policies. 
Distr-Battery has $13\%$ additional costs due to the extra battery capacity. 
Centr-Global and Centr-Graph have lower server costs since they need fewer deployed servers 
than \name{} to reach the same computing capacity. As the centralized data centers aggregate the power supply 
from multiple renewable energy farms, which generates a smoothened power supply curve, allowing Centr-Global and Centr-Graph 
to provision fewer servers to make use of their peak power supply.
However, Centr-Global and Centr-Graph have to connect all renewable energy farms to the centralized \siteName{} via 
power transmission lines,  
causing higher cost ($62\times$ more than \name{}). Thus, these solutions are less practical 
in terms of cost efficiency.

\begin{figure}[t!]
  \centering
    \includegraphics[scale=0.88]{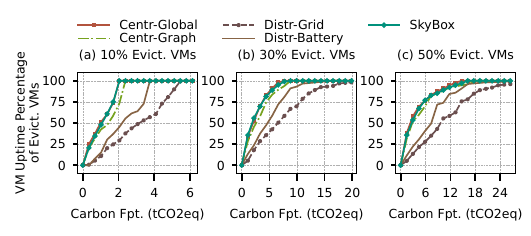}
    \vspace{-3ex}
	\caption{Efficiency of \siteName{s} with different percentages of evictable VMs (quantified with evictable VM uptime).}
  \label{fig:eval:performance}
  \vspace{-3.5ex}
\end{figure}

\begin{figure}[t]
  \centering
  \includegraphics[scale=0.95]{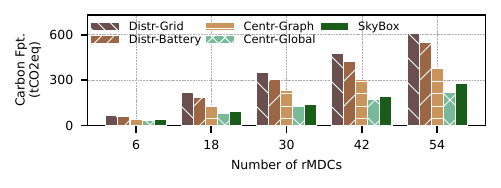}
	 \vspace{-3ex}
	 \caption{Total carbon footprint with different number of sites.}
  \label{fig:eval:scale}
	 \vspace{-3ex}
\end{figure}

\begin{figure}[t]
  \centering
  \includegraphics[scale=0.95]{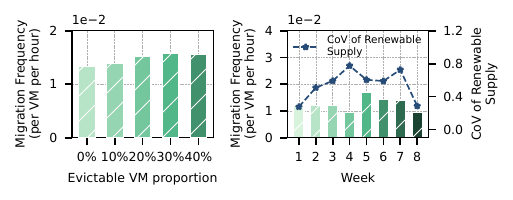}
	 \vspace{-3ex}
	 \caption{VM migration frequency with different VM settings.}
  \label{fig:eval:migr_freq}
	 \vspace{-3ex}
\end{figure}

\subsection{VM Uptime Improvement}
\label{sec:eval:downtime}

\name{} improves the overall efficiency of \siteName{s} as reflected by the evictable VM uptime.  
Figure~\ref{fig:eval:performance} presents the tradeoff between evictable VM uptime and their operational carbon footprint, with different settings of the evictable VMs (from $10\%$ to $50\%$). 
With the same operational carbon footprint, \name{} achieves the similar efficiency as Centr-Global and Centr-Graph. This is because they have relatively 
stable renewable energy supplies, which minimizes the negative impact (i.e., VM shutdown) on evictable VMs. 

\name{} always delivers better efficiency than Distr-Grid and Distr-Battery. For instance, Figure~\ref{fig:eval:performance} (left) shows that, when the carbon footprint is $2.0$ tCO2eq, 
\name{} ensures the evictable VMs have $100\%$ uptime, while Distr-Grid and Distr-Battery only have $29\%$ and $45\%$ uptime, respectively. 
This is because \name{} can timely migrate VMs from the \siteName{s} that have insufficient renewable power, and keep these VMs operational.  
But Distr-Grid and Distr-Battery have to suspend these VMs, otherwise, they have to use power grid, resulting in increased operational 
carbon footprint. 

\name{} does not rely on evictable VMs to gain benefits. 
With different proportions of evictable VMs (10\%-50\%), even when the uptime percentage of all evictable VMs keeps at $100\%$, 
\name{} still effectively reduces the carbon footprint, because of the stable renewable power offered 
in subgraphs.

\subsection{Sensitivity Analysis}
\label{sec:eval:sensitivity}

\noindent
\textbf{Varying the number of \siteName{s}:}
We show the scalability of \name{} by increasing the number of \siteName{s} from 6 to 54.
As shown in Figure \ref{fig:eval:scale}, the total carbon footprint of \name{} is close to ideal, 
and \name{} consistently outperforms Distr-Grid and Distr-Battery as we increase the number of \siteName{s}.
We also observe that \name{} has even greater carbon reduction (from $45\%$ to $56\%$) with more \siteName{s}.
This is because a larger pool of candidate subgraphs (e.g., 20 candidates with 6 sites vs. 24.8K candidates 
with 54 sites) increases the opportunities for \name{} to identify more stable subgraphs.

\noindent  
\textbf{Varying migration frequency:} 
We evaluate \name{} under different migration frequencies, where the frequency is measured by the number of migration events per VM per hour of runtime. 
Figure~\ref{fig:eval:migr_freq} (left) shows that under different evictable VM proportions, \name{} obtains consistent carbon footprint reduction 
with low migration frequency (around $0.015$). 
Consider that live migration has low power consumption, only tens of milliseconds of VM downtime, and low impact on application performance~\cite{freelunch, clark2005live, vm_migration_hpdc11}, 
\name{} does not incur much migration overhead. We show that \name{} delivers constant migration frequency over time (Figure~\ref{fig:eval:migr_freq}, right), under the setting 
of 0\% evictable VMs.

\begin{figure}[t]
  \centering
  \includegraphics[scale=0.91]{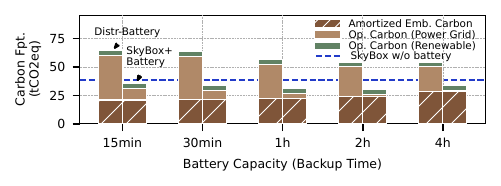}
	 \vspace{-3ex}
	 \caption{Total carbon footprint with various battery capacities.}
  \label{fig:eval:battery}
	 \vspace{-3ex}
\end{figure}

\noindent
\textbf{Varying battery capacity:}
We present the benefit of \name{} using different battery capacities (from a 15-minute to a 4-hour backup time). 
As shown in Figure~\ref{fig:eval:battery}, the increased battery capacities help offset more operational carbon footprint for \name{}. 
\name{} without extra battery can bring more benefits than Distr-Battery with a large battery capacity.
This is because Distr-Battery has limited opportunity to keep the batteries charged upon insufficient renewable power supply, while \name{} can migrate VMs to another \siteName{}.

\begin{figure}[t]
  \centering
  \includegraphics[scale=0.91]{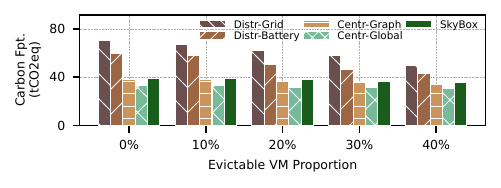}
	 \vspace{-3ex}
	 \caption{Total carbon cost with various evictable VMs.}
  \label{fig:eval:low_priority_ratio}
	 \vspace{-3ex}
\end{figure}

\noindent
\textbf{Varying evictable VM proportion:}
In Figure~\ref{fig:eval:low_priority_ratio}, we change the proportion of the evictable VMs from $0\%$ to $40\%$. The total carbon footprint decreases 
for all the placement policies, as evictable VMs offer more flexibility in the VM scheduling. \name{} consistently outperforms Distr-Grid and Distr-Battery. 
Even with $0\%$ evictable VMs (i.e., all VMs are regular), \name{} outperform Distr-Grid and Distr-Battery, with a reduction of the total carbon footprint of $44\%$ and $35\%$ respectively. 
\name{} also incurs only $2\%$ and $14\%$ extra carbon compared to the Centr-Graph and Centr-Global, respectively.

\subsection{Benefit Breakdown of \name{}}
\label{sec:eval:breakdown}

\begin{figure}[t]
  \centering
  \includegraphics[width=0.42\textwidth]{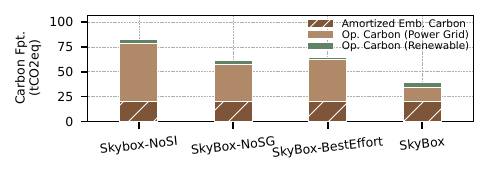}
	 \vspace{-3ex}
	 \caption{Benefits of different \name{} components.}
  \label{fig:eval:step}
	 \vspace{-3ex}
\end{figure}

We now evaluate the benefit of each \name{} component. We compare \name{} with its variants, 
each replacing one \name{} component with a baseline policy (e.g., MIP Model vs. Best-Effort VM allocation).
We first show the comparison of the total carbon footprint in Figure~\ref{fig:eval:step}. 
By identifying stabler renewable energy farms, \name{} reduces the total carbon footprint by $53\%$, compared to the random site selection scheme (\name{-NoSI}). 
With subgraph identification, \name{} groups multiple \siteName{s} to further achieve stable aggregated power production. It helps \name{} to reduce $37\%$ 
carbon footprint compared with the random subgraph identification (\name{-NoSG}). Furthermore, \name{} can find a more optimized VM placement and migration 
plan using our proposed MIP model, leading to a $39\%$ less carbon footprint than the best-effort VM placement policy (\name{-BestEffort}). 

\begin{figure}[t]
\centering
\includegraphics[width=0.44\textwidth,trim={0 0cm 0 0cm}]{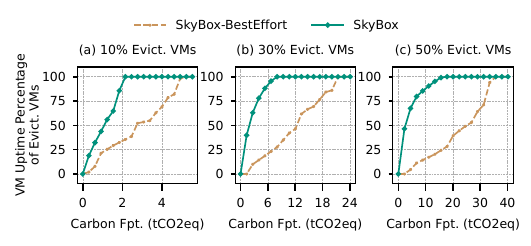}
\vspace{-3ex}
	\caption{Comparison of the VM placement/migration policy in \name{} vs. the best-effort approach.}
\label{fig:eval:performance_b}
\vspace{-3.5ex}
\end{figure}

To further evaluate how the MIP model of \name{} optimizes VM placement and migration, we compare \name{} against \name{-BestEffort}.
In Figure~\ref{fig:eval:performance_b}, with the same carbon footprint, \name{} achieves higher VM uptime for evictable VMs. 
Their gap is larger with more evictable VMs. This is because the scheduling of a larger set of VMs across \siteName{s} forms 
an more complicated decision space, it is harder for the heuristic-based approach (i.e., best-effort) to 
find optimized VM placement and migration plans.  

\subsection{Resilience to Mispredictions}
\label{sec:eval:misprediction}

\begin{figure}[t]
  \centering
  \includegraphics[width=0.45\textwidth]{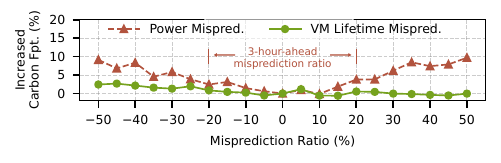}
	 \vspace{-2ex}
	 \caption{The impact of different misprediction ratios of power supply and VM lifetime in \name{}.}
  \label{fig:eval:power_mispred}
   \vspace{-4ex}
\end{figure}

We now demonstrate the robustness of \name{} in handling the mispredictions of power supply and VM lifetime. 
In Figure~\ref{fig:eval:power_mispred}, we vary the maximum misprediction ratios of the renewable power production 
and the VM lifetime from $-50\%$ to $50\%$, and present the percentage of increased carbon footprint against \name{} with accurate predictions.  

The mispredictions of power production incur a small increase in the total carbon footprint (0.6\%-9.7\%). 
And in practice, power production usually has a low prediction error. For instance, the misprediction ratio of 
a 3-hour-ahead power production prediction is typically $\pm20\%$ ($9\%$ on average, see $\S$\ref{sec:predictability}), 
where \name{} has less than $3.7\%$ carbon footprint overhead.
As for VM lifetime mispredictions, their impact is trivial (up to $3\%$) in \name{}. 


\section{Related Work}
\label{sec:related}

\noindent\textbf{Renewable energy for data centers.}
Using renewable energy to power data centers has been investigated in prior 
studies~\cite{parasol:asplos2013, greenhadoop, 
inigo:micro2014, goiri2011greenslot, stewart2009somejoules, iswitch:isca2012, chao:micro2013, ren2012carbon}. 
They benefited the recent development of containerized, mobile, truck, and edge data centers~\cite{mobiledc,containerdc, truckdc} 
that enable platform operators to move data centers closer to renewable energy farms with low cost. 
Most recently, 
a majority of popular cloud providers have been planning to use renewable energy in their 
data centers~\cite{google:energy, azurepledge, carbonexplorer:asplos2023, aws:energy}, and  
discussed how to build carbon-aware data centers with renewable energy~\cite{carbonexplorer:asplos2023}. Our work \name{} shares the same goal with these prior studies, 
but proposes a new approach of building modular data centers at scale.

\noindent\textbf{Energy and carbon efficiency of data centers.}
To improve the energy efficiency of data centers, a variety of computer architecture and systems techniques 
have been developed~\cite{david2010rapl, kontorinis2012managing, greenai, subramanian2010reducing, 
wu2014green, oversubscript, david2011memory}, including the dynamic voltage and frequency scaling, and the 
power management in different server components (e.g., CPU cores, caches, and memory). Researchers also 
developed many power-aware scheduling policies across the entire systems stack~\cite{von2009power, chatzipapas2015challenge, 
protean, borg, flex, energyaware, wang2013data, qoop, buttazzo2002elastic, zcc, greenhadoop, goiri2011greenslot, 
greenswitch, freelunch}, such as workload scheduling and VM placement. 
Recently, more studies have focused on improving the carbon efficiency~\cite{gupta2021chasing, gupta2022act, wiesner2021let, souza2023ecovisor, radovanovic2022carbon, irwin-api, wang2023peeling}. 
These studies are orthogonal to our work. 
\name{} focuses on exploring the efficiency of renewable energy with 
modular data centers.

\noindent\textbf{Energy-based elasticity and reliability.}
To provide reliable and elastic computing for cloud services, prior studies proposed power-aware workload migration 
policies across data centers, depending on the energy cost and availability~\cite{liu2014pricing,bahrami2018data,chen2014data,
liu2013data}. And they investigated the use of VM migration~\cite{spotcheck,yank,ruprecht2018vm}, 
replication~\cite{cully2008remus,lagar2009snowflock}, and checkpointing~\cite{scr-checkpoint, dnn-checkpoint} 
to manage faults and power outages. Unlike these prior studies, \name{} develops new algorithms for VM placement and migrations across geo-distributed modular data centers.

\section{Conclusion}
\label{sec:conclusion}
We present \name{}, a framework for facilitating the deployment of geo-distributed 
modular data centers at scale. It tackles the power variability of renewables with the insights derived 
from the study of real-world power traces. \name{} presents a set of techniques to assist datacenter vendors to 
identify the suitable sites for deploying modular data centers, and enable efficient VM placement and migrations. 

\bibliographystyle{IEEEtranS}
\bibliography{skybox}

\end{document}